\documentclass[fleqn,10pt]{wlscirep}

\newcommand{\E}{\mathcal{E}}
\newcommand{\G}{\mbox{Geo}}
\newcommand{\F}{\mathcal{F}}

\title{Evolutionary dynamics of incubation periods }

\author[a]{Bertrand Ottino-L\"{o}ffler}
\author[b]{Jacob G. Scott} 
\author[a,*]{Steven H. Strogatz}
\affil[a]{Center for Applied Mathematics, Cornell University, Ithaca, New York 14853, USA}
\affil[b]{Departments of Translational Hematology and Oncology Research and Radiation Oncology, Cleveland Clinic, Cleveland, Ohio 44195, USA}
\affil[*]{strogatz@cornell.edu}


\begin{abstract} 

The incubation period of a disease is the time between an initiating pathologic event and the onset of symptoms~\cite{armenian1983incubation}. For typhoid fever~\cite{sawyer1914typhoid,miner1922incubation}, polio~\cite{sartwell1952incubation}, measles~\cite{goodall1931incubation}, leukemia~\cite{feinleib60} and many other diseases~\cite{boag49,sartwell1950distribution,sartwell66,lessler09}, the incubation period is highly variable. Some affected people take much longer than average to show symptoms, leading to a distribution of incubation periods that is right skewed and often approximately lognormal~\cite{sartwell1950distribution,sartwell66,lessler09}. Although this statistical pattern was discovered more than sixty years ago~\cite{sartwell1950distribution}, it remains an open question to explain its ubiquity~\cite{nishiura07}. Here we propose an explanation based on evolutionary dynamics on graphs~\cite{moran58, williams72,lieberman05,nowak2006evolutionary,ohtsuki2006simple,frean13,ashcroft15}. For simple models of a mutant or pathogen invading a network-structured population of healthy cells, we show that skewed distributions of incubation periods emerge for a wide range of assumptions about invader fitness, competition dynamics, and network structure. The skewness stems from stochastic mechanisms associated with two classic problems in probability theory: the coupon collector and the random walk~\cite{posfai10,feller1968introduction}. Unlike previous explanations~\cite{nishiura07,horner1992criteria} that rely crucially on heterogeneity, our results hold even for homogeneous populations. Thus, we predict that two equally healthy individuals subjected to equal doses of equally pathogenic agents may, by chance alone, show remarkably different time courses of disease. 
\end{abstract}

\date{\today}

\begin{document}
\flushbottom
\maketitle

\begin{figure}[hb] 
\centering
\includegraphics[width = 0.78\textwidth]{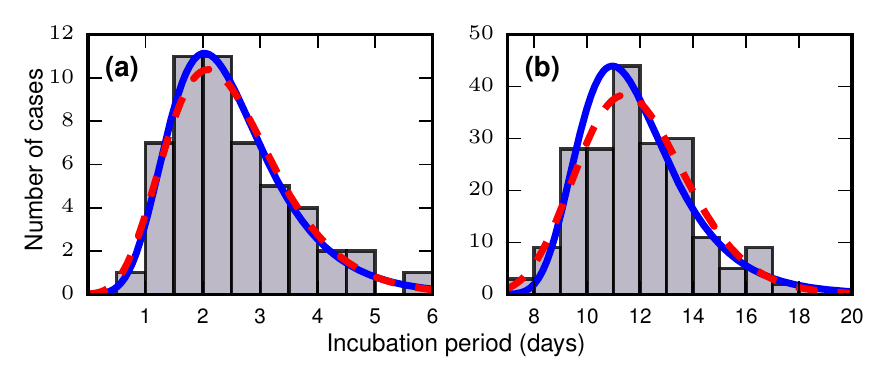} 
\caption{\textbf{Frequency distributions of incubation periods for two infectious diseases.} Data redrawn from historic examples. Dashed red curves are noncentral lognormal distributions. Solid blue curves are Gumbel distributions, predicted by the theory developed here. Both sets of curves were fitted via the method of moments. (a) Data from an outbreak of food-borne streptococcal sore throat, reported in 1950 ~\cite{sartwell1950distribution}. (b) Data from the 1940-41 measles epidemic in New York City, reported in 1944~\cite{stillerman1944attack}. }
\label{historic}
\end{figure}

The discovery that incubation periods tend to follow right-skewed distributions originally came from epidemiological investigations of incidents in which many people were simultaneously and inadvertently exposed to a pathogen. For example, at a church dinner in Hanford, California on March 17, 1914, ninety-three individuals became infected with typhoid fever after eating contaminated spaghetti prepared by an asymptomatic carrier known to posterity as Mrs. X. Using the known time of exposure and onset of symptoms for the 93 cases, Sawyer~\cite{sawyer1914typhoid} found that the incubation periods ranged from 3 to 29 days, with a mode of only 6 days and a distribution that was strongly skewed to the right. Similar results were later found for other infectious diseases. Surveying the literature in 1950, Sartwell noted a striking  pattern: the incubation periods of diseases as diverse as streptococcal sore throat~\cite{sartwell1950distribution} (Fig.~\ref{historic}a),  measles~\cite{stillerman1944attack} (Fig.~\ref{historic}b), polio, malaria, chicken pox, and the common cold were all, to a good approximation, lognormally distributed~\cite{sartwell1950distribution}.

Two natural questions arise: Why should incubation periods be distributed at all, and why should they be distributed in the same way for different diseases? Previous explanations rest on the presumed heterogeneity of the host, the pathogen, or the dose~\cite{sartwell1950distribution, nishiura07,horner1992criteria}. To see how this works, return to the typhoid outbreak at the Hanford church dinner~\cite{sawyer1914typhoid}. Every person who ate that spaghetti presumably had a different level of overall health and immune function, and every plate of spaghetti was likely contaminated with a different dose and possibly even strain of typhoid. 
Suppose the typhoid bacteria proliferated exponentially fast within the hosts  and triggered symptoms when they reached a fixed threshold. Then if the bacterial dose, growth rate, or triggering threshold were normally distributed across the hosts, one can show that the resulting distribution of incubation periods would have been either exactly or approximately lognormal (see Methods, ``Influence of heterogeneity''). On the other hand, there is counter-evidence that lognormal distributions can occur even if some of these sources of heterogeneity are lacking. For example,  Sartwell~\cite{sartwell1950distribution} reanalyzed data from a study~\cite{bodian49} in which identical doses and strains of polio virus were injected into the brains of hundreds of rhesus monkeys. The incubation period, defined as the time from inoculation to the onset of paralysis, was still found to be approximately lognormally distributed, even though the route of infection and the viral dose and strain were held constant. Moreover, the lognormal distributions commonly observed for human diseases have a particular shape, with a dispersion factor~\cite{sartwell1950distribution} around $1.1-1.5$, which previous models cannot explain without special parameter tuning.

Here we propose a new explanation for the skewed distribution of incubation periods. Instead of heterogeneity, it relies on the stochastic dynamics of the incubation process, as the pathogen invades, multiplies, and competes with itself and the cells of the host. The theory predicts that under a broad range of circumstances, incubation periods should follow a right-skewed distribution that resembles a lognormal, but is actually a Gumbel, one of the universal extreme value distributions~\cite{kotz2000extreme}.  Heterogeneity is not required, but it is allowed; it does not qualitatively alter our results when included. 

\begin{figure}[t] 
\includegraphics[width = 0.6\textwidth]{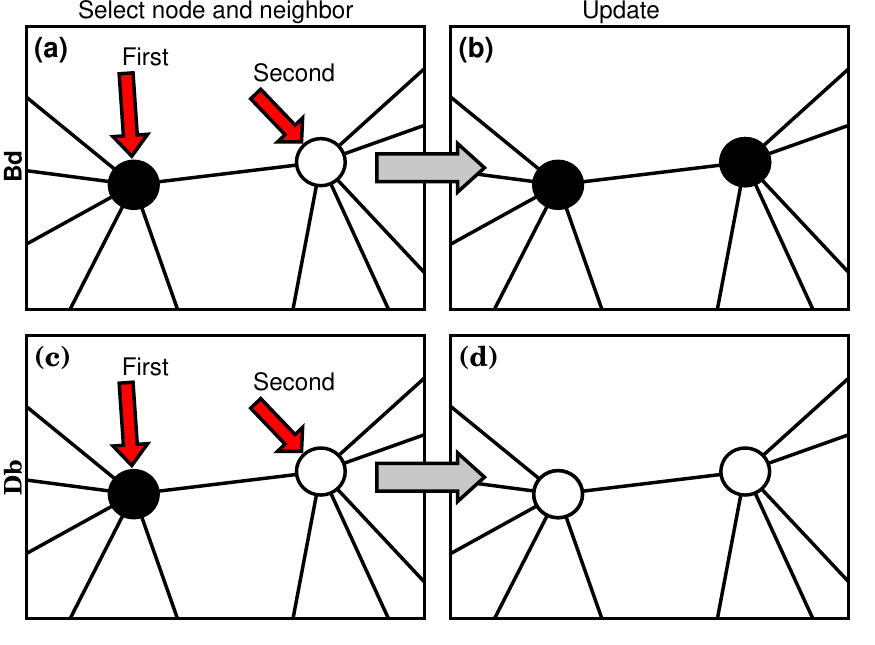}
\caption{\textbf{Evolutionary update rules.} (a) In the Birth-death (Bd) update rule, a node anywhere in the network is selected at random, with probability proportional to its fitness, and one of its neighbors is selected at random, uniformly. (b) The neighbor takes on the type of the first node. In biological terms, one can interpret this rule in two ways: either the first node transforms the second; or it gives birth to an identical offspring that replaces the second. (c) In the Death-birth (Db) update rule, a node is selected at random to die, with probability inversely proportional to its fitness, and one of its neighbors is selected at random, uniformly, to give birth to one offspring. (d) The first node is replaced by the offspring of the second.}
\label{plot_schematic}
\end{figure}

We model the incubation process using the formalism of evolutionary graph theory~\cite{lieberman05,nowak2006evolutionary,ohtsuki2006simple,ashcroft15}. A network of $N \gg 1$ nodes is used to represent an environment within a host where a pathogen is invading and reproducing. For example, the network could represent the intestinal microbiome, where harmful typhoid bacteria are competing against a benign resident population of gut flora in a mixing system (modeled as a complete graph); or it could represent mutated leukemia cells vying for space against healthy hematopoietic stem cells within the well-organized three-dimensional bone marrow space (modeled as a 3D lattice); or it could represent the flat tracheal epithelium with influenza sequentially compromising nearby healthy cells (modeled as a 2D lattice). For the sake of generality, we will refer to the two types of agents as healthy residents and harmful invaders.

The invasion dynamics are modeled as a Moran process~\cite{moran58,williams72,lieberman05,nowak2006evolutionary}. Invaders are assigned a relative fitness $r$ (suggestively called the carcinogenic advantage by Williams and Bjerknes~\cite{williams72}). The fitness of residents is normalized to 1. We consider two versions of the Moran process. In the Birth-death (Bd) version (Fig.~\ref{plot_schematic}a), a random node is chosen, with probability proportional to its fitness. It gives birth to a single offspring. Then one of its neighbors is chosen uniformly at random to die and is replaced by the offspring (Fig.~\ref{plot_schematic}b). We also consider Death-birth (Db) updates  (Figs.~\ref{plot_schematic}c, d). In this version of the model, a node is randomly selected for death, with probability proportional to $1/r$; then a copy of a uniformly random neighbor replaces it. To test the robustness of our  results, we study both versions of the Moran model on various networks: complete graphs, star graphs, Erd\H{o}s-R\'{e}nyi random graphs, one-, two-, and three-dimensional lattices, and small-world, scale-free, and $k$-regular networks. We also vary the invader fitness $r$ and the model criterion for the onset of symptoms. These extensions are presented in the Methods, Extended Data Figs.~\ref{plot_InfBd}-\ref{plot_truncs_and_fits}, and Supplementary Information. Here we focus on the simplest cases to expose the basic mechanisms.

Our simulations start with a single invader placed at a random node in a network of otherwise healthy residents. The update rule is applied at discrete time steps. In the long run, either the invaders replace all the residents, or vice versa. Supposing that symptoms are triggered when the entire network has been taken over by invaders, then the incubation period is the number of time steps between the introduction of the invader and its fixation. On the other hand, if the invaders die out and the healthy cells take over, then the process is stopped and no observable symptoms manifest.

\begin{figure}[ht] 
\includegraphics[width = 0.8\textwidth]{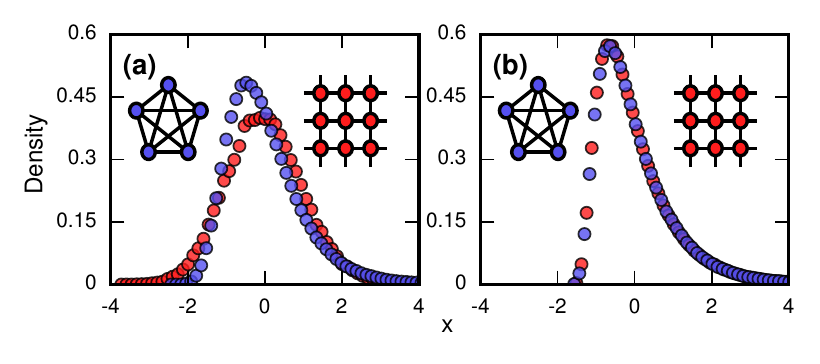} 
\caption{\textbf{Network topology and invader fitness shape the distribution of incubation periods.} Plots show simulated distributions of incubation periods, defined here as invader fixation times. Starting from a single invader at a random node, the state of the network was updated by Birth-death dynamics on both a complete graph and a two-dimensional (2D) lattice. Results for the Death-birth update rule (not shown) are identical.  All distributions are normalized to have zero mean and unit variance. (a)  Infinitely fit invader. For invader fitness $r \rightarrow \infty$, the distribution is right-skewed for a complete graph (blue symbols). It approaches a Gumbel distribution as $N \rightarrow \infty$, where $N$ is the number of nodes in the network. In contrast, for a 2D lattice (red symbols) the incubation periods are normally distributed. The difference is that a coupon collection mechanism operates in the complete graph and in lattices of sufficiently high dimension $d \ge 3$; this mechanism causes the right skew. Simulations used $10^6$ repetitions on a complete graph of $N = 150$ nodes, and $10^5$ repetitions for a 2D lattice of $N = 30^2$ nodes. (b) Neutrally fit invader.  Distributions of incubation periods are shown for invader fitness $r=1$, using $10^6$ repetitions on a complete graph of $N = 50$ nodes (blue symbols), and $10^5$ repetitions for a 2D lattice of $N = 7^2$ nodes (red symbols). Similar right-skewed distributions occur for both networks, caused by a  conditioned random walk mechanism. }
\label{plot_MultiR}
\end{figure}

First consider what happens if the invaders have infinite fitness ($r \rightarrow \infty$). While an exaggeration, this case is instructive and is a reasonable approximation for aggressive cancerous mutations or certain viral infections. In this limit, the dynamics simplify enormously: only the invaders reproduce. But because they give birth and replace their neighbors blindly, they waste time steps whenever they compete between themselves and one invader replaces another. These random self-replacements slow down the incubation process, and make it highly variable.  In fact, the level of in-fighting is what determines the incubation period in this case. Beyond fitness, the topology of the network matters too. For low-dimensional networks, exemplified by a two-dimensional  lattice (Figure~\ref{plot_MultiR}a, red circles), the growth rate of the invader population remains roughly constant as takeover occurs. This leads to a normal distribution of incubation periods  (Figure~\ref{plot_MultiR}a, red circles; and see Methods, ``Birth-death, other solvable networks''). However, on very high-dimensional networks like the complete graph (Fig.~\ref{plot_MultiR}a, blue circles), the distribution becomes right skewed. Intuitively, this happens because every invader now has a chance of replacing any healthy node or any other invader. It is as if at every time step a node  gets blindly drawn from a bag, relabeled as an invader, and returned to the bag. At the start of the incubation process, almost every draw adds another invader to the population and the infection progresses rapidly. But near the end, it will take many, many draws to blindly fish out the last remaining healthy node, as needed to terminate the incubation period. This slowing-down phenomenon near the end should feel familiar to anyone who has tried to complete a collection of baseball cards, stamps, or coupons, since they are all manifestations of the coupon collector’s problem, a well-studied concept in probability theory~\cite{posfai10,feller1968introduction, erdos61}. Because of those frustratingly long waits to collect the final healthy node, the incubation period distribution gets skewed to the right. In the infinite-$N$ limit (see Methods, ``Birth-death, complete graph''), the coupon collector's process returns a Gumbel distribution, which resembles a lognormal and can be mistaken for it~\cite{read98}. Indeed, when a Gumbel and a lognormal are fit to the same real data, as in Fig.~\ref{historic}, it is hard to tell them apart. 

At the other extreme, suppose the invaders have no selective advantage ($r=1$). Then a different stochastic mechanism skews the distribution of incubation periods to the right  (Fig.~\ref{plot_MultiR}b and Methods, ``Random Walk Skewness''). For many networks, the dynamics reduce to an unbiased random walk on the number of invaders, with waiting times at each population level. There are two absorbing states, corresponding to both $0$ and $N$ invaders for the two kinds of fixation. However, we only care about random walks that successfully hit $N$, as these represent disease processes that manifest symptoms, so we must always condition on its success. This demands that the invader experience early success and growth, pushing it away from probable extinction. This conditioning introduces a bias which makes short incubation times probable, but long walks may still occasionally occur, driving the mean time above the median. In short, a conditioned random walk will introduce a right skew in the distribution of incubation periods. This effect holds for both high- and low-dimensional networks (Fig.~\ref{plot_MultiR}b), and for Birth-death and Death-birth dynamics. 

Thus, our model suggests two basic mechanisms underlying the observed right-skewed, approximately lognormal distributions of incubation periods. When the fitness of the pathogen is high, the skew comes from coupon collection; when the pathogen fitness is neutral or low, the skew comes from conditioned random walks. Neither of these effects demand any heterogeneity from the invader or the host. However, the model can accommodate such heterogeneity, either by having the invader fitness $r$ be randomly drawn, or by having symptoms occur when a random fraction $f$ of the host network has been invaded. Our simulations show that both sources of heterogeneity only exaggerate the level of right-skewness we would have seen without them (Methods, ``Influence of heterogeneity" and Extended Data Fig.~\ref{plot_truncs_and_fits}).  

Beyond accounting qualitatively for the distributions of incubation periods, our model accounts for a quantitative feature that has never been explained before.  As shown in Extended Data Table~\ref{table_Disp}, the distributions generated by highly fit pathogens and mutants are predicted to have dispersion factors (also known as geometric standard deviations) of about $1.1-1.4$, close to the actual values of $1.1-1.5$ observed for various  diseases~\cite{sartwell1950distribution,sartwell66,nishiura07}. The model also helps to explain why so few diseases yield dispersion factors greater than 1.5. Such high dispersion factors arise only for $r \approx 1$, corresponding to pathogens or mutants that are only slightly more fit than the resident populations against which they are competing.   

In 1546, Fracastorii~\cite{fracastorii} described the incubation of rabies after a bite from an rabid dog as ``stealthy, slow, and gradual." Today, nearly five centuries later, the dynamics of incubation processes remain stealthy and slow to yield their secrets. We have tried to shed light on their patterns of variability with the help of a new conceptual tool, evolutionary graph theory. This approach provides a possible solution to the longstanding question of why so many disparate diseases show such similarly-shaped distributions of incubation periods. What remains is to quantify the dynamics of incubation processes experimentally with high-resolution measurements in time and space. 

\bigskip



\section*{Methods and Extended Data Figures and Table}
Here we describe the model and our analytical and numerical results in further detail. We also test the robustness of our claims with respect to relaxation of the various assumptions in the model. See Supplementary Information for complete proofs of analytical results. No experiments were conducted, and no original empirical data were collected. 

\subsection*{Birth-death, complete graph}

The population of cells is represented by a network of $N$ nodes. Edges between nodes indicate which cells can potentially interact with each other.  There are two types of cells: harmful invaders with fitness $r$, and healthy residents with fitness 1. All simulations are initialized with a single invader placed at a random node. 

The Moran Birth-death (Bd) update rule has two steps. First, a node is randomly selected out of the total population, with probability proportional to its fitness. Second, a neighbor of the first node is chosen, uniformly at random, and takes on the type of the first node. 

In a complete graph, all nodes are adjacent. Therefore, the probability of adding a new invader, given there are currently $m$ invaders, is

\begin{equation*}
p_m := P(\mbox{Choose an invader}) \cdot P(\mbox{Neighbor is resident}) = \frac{mr}{mr + (N-m)} \cdot \frac{N-m}{N-1} .
\end{equation*}

\noindent In the infinite fitness limit $(r\rightarrow \infty)$, the first term approaches 1 and we get

\begin{equation*}
p_m := \frac{N-m}{N-1},
\end{equation*}

\noindent and the probability of the invader population ever decreasing is 0. So the time $T$ to invader fixation is sum of all the transition times $m \to m+1$ for $m = 1, 2, \ldots, N-1$. These transition times can be calculated as follows. For the population to take $t$ steps to go from $m$ to $m+1$ invaders, nothing must have happened for $t-1$ steps before advancing on the $t$'th step. The probability of this happening is exactly

\begin{equation*}
p_m(1-p_m)^{t-1}.
\end{equation*}

\noindent In other words, the time to add a new invader is exactly a geometric random variable. Therefore, the total fixation time is just

\begin{equation*}
T = \sum_{m=1}^{N-1} \mbox{Geo}(p_m) = \sum_{k=1}^{N-1} \mbox{Geo}\left(\frac{k}{N-1}\right).
\end{equation*}

\noindent This random variable $T$ describes a process identical to that of the coupon collector's problem~\cite{posfai10,feller1968introduction}. In both, we have a collection of $N-1$ nodes, and draw a random one with replacement at each time step. If we pick a healthy node, we relabel it and toss it back, and repeat until there are no healthy nodes left. By adapting classic results~\cite{erdos61,baum65}, we show in the Supplementary Information that it is straightforward to find the asymptotic distribution of $T$ as $N$ gets large. 
To normalize this distribution, note that its mean is $\mu = \sum_m p_{m}^{-1} \approx N\log(N) + N\gamma$. Then we find 

\begin{equation}\label{Bd_Complete}
\frac{T - \mu}{N} \xrightarrow{d} \mbox{Gumbel}(-\gamma, 1).
\end{equation}

\noindent Here $\gamma \approx 0.5772$ is the Euler-Mascheroni constant, $\xrightarrow{d}$ denotes convergence in distribution, and a Gumbel($\alpha, \beta$) random variable has a density given by

\begin{equation}\label{GumbelDist}
h(x) = \beta^{-1} e^{-(x-\alpha)/\beta}\exp\left(- e^{-(x-\alpha)/\beta} \right).
\end{equation}

\noindent This prediction for the normalized distribution of the incubation period $T$ agrees with simulations on large networks  (Fig.~\ref{plot_InfBd}a). 

A Gumbel distribution of incubation periods has previously been obtained for a variant of this model. Instead of working with the large-$N$ limit of a complete graph, it assumed a continuous-time birth-death model of an invading microbial population whose dynamics were governed by differential equations~\cite{williams65}.

\subsection*{Birth-death, other solvable networks}

The analysis of the finite-$N$ complete graph sets up an important framework that  can be applied to more complicated networks. For example, in the Supplementary Information we prove that the distribution of fixation times $T$ for a star network also converges to a Gumbel for $N \gg 1$. Specifically, 

\begin{equation}\label{Bd_Star}
\frac{T - N^2\log(N) - (\gamma-1)N^2}{N^2} \xrightarrow{d} \mbox{Gumbel}(-\gamma, 1).
\end{equation}

\begin{figure*}[ht] 
\includegraphics[width = 0.6\textwidth]{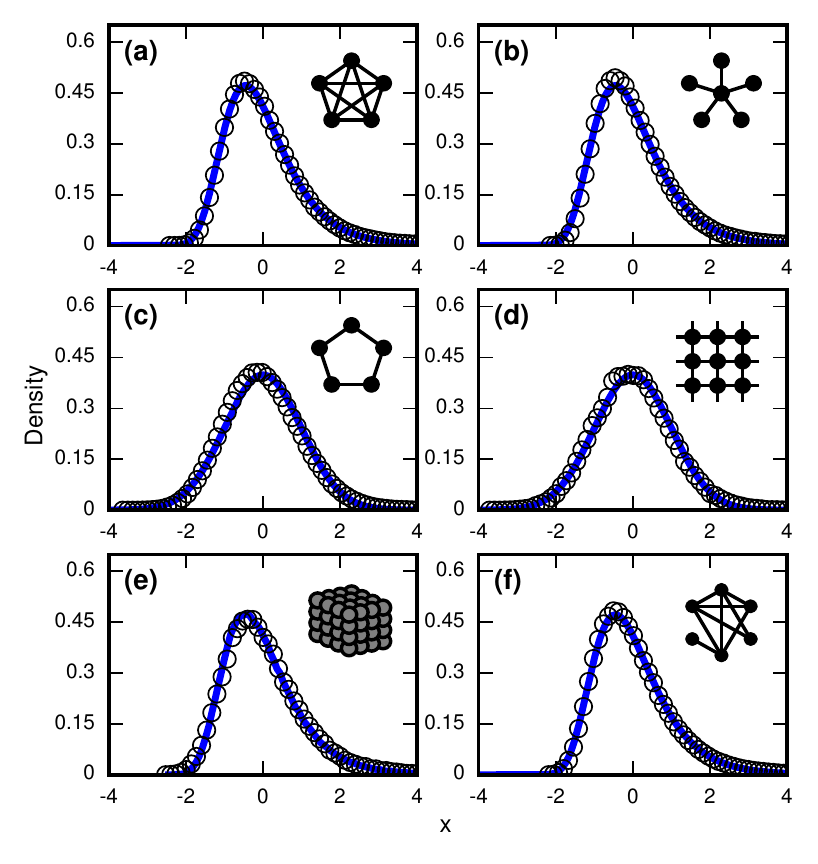} 
\caption{\textbf{Network dependence of incubation periods.} Distributions of invader fixation times, normalized by their means and standard deviations, are shown for infinite-$r$ Birth-death dynamics on various networks. Open circles show simulation results. Curves show analytical predictions. Insets show schematics of networks. (a) The distribution of fixation times for a complete graph on $N=150$ nodes, for $10^6$ runs. Mean $\mu$ and standard deviation $\sigma$ were calculated analytically. Curve shows a Gumbel distribution. (b) The Gumbel distribution of fixation times for a star graph with $N=75$ spokes, for $10^6$ runs. $\mu$ and $\sigma$ were calculated analytically. (c) Normal distribution of fixation times for a 1D ring on $N=75$ nodes, for $10^6$ runs. $\mu$ and $\sigma$ were calculated analytically. (d) Normal  distribution of fixation times for a 2D lattice of $N = 60 \times 60$ nodes, for  $10^5$ runs. Here $\mu$ and $\sigma$ were calculated empirically. (e) The distribution of fixation times for a 3D lattice of $N = 11^3$ nodes, for $10^5$ runs. $\mu$ and $\sigma$ were calculated empirically. The predicted distribution is the result of an approximating sum of exponential random variables under $10^6$ repetitions. (f) The distribution of fixation times for an Erd\H{o}s-R\'{e}nyi random graph on $N = 115$ nodes with an edge probability of $\rho = 0.5$. $\mu$ and $\sigma$ were calculated empirically. }
\label{plot_InfBd}
\end{figure*}

\noindent This prediction matches simulations (Fig.~\ref{plot_InfBd}b).

The same framework also applies to a one-dimensional (1D) ring lattice, but instead of using the coupon-collector framework, we need to cite the Lindeberg-Feller central limit theorem~\cite{durrett91}. As shown in the Supplementary Information, this gives us
\begin{equation}\label{Bd_Normal}
\frac{T - (N^2 - N)/2}{(2N^3 -3N^2 + N)/6}\xrightarrow{d} \mbox{Normal}(0, 1).
\end{equation}

\noindent This prediction agrees with simulations (Fig.~\ref{plot_InfBd}c). 

For a two-dimensional square lattice, it is more difficult to produce analytical results that are both rigorous and exact. But by making an approximation based on the geometry of the lattice (see Supplementary Information), we can make a non-rigorous analytical guess about the distribution of the fixation times $T$. Via these arguments, and given $\mu = \mbox{E}[T]$ and $\sigma^2 = \mbox{Var}(T)$, we predict  

\begin{equation}\label{Bd_2DLattice}
\frac{T - \mu}{\sigma} \xrightarrow{d} \mbox{Normal}(0, 1).
\end{equation}

\noindent Despite the approximation, this prediction works well (Fig.~\ref{plot_InfBd}d). 

By similar arguments, we predict that lattices of dimension $d \ge 3$  have right-skewed asymptotic distributions of fixation times.  Specifically, given $\eta := 1 - 1/d$, we predict

\begin{equation}\label{Bd_LatticeSkew}
\mbox{Skew}(T) := \frac{E[(T-\mu)^3]}{\sigma^3} = \frac{2 \zeta(3\eta)}{\zeta(2\eta)^{3/2}},
\end{equation}

\noindent where $\zeta$ is the Riemann zeta function. The methods used to derive that can also be used to create approximate finite-size distributions for the lattices (Fig.~\ref{plot_InfBd}e).

In particular, we predict positive skew for all $d\geq3$ and for the skew to increase monotonically with dimension (Supplementary Information). Meanwhile, both 1D and 2D lattices have normal asymptotic distributions, and therefore no skew. This establishes $d=2$ as a critical dimension in these dynamics, transitioning from zero skew to positive skew. 

Incidentally, these arguments also suggest that appropriate infinite-dimensional networks will asymptotically have a Gumbel distribution. This is numerically true for the Erd\H{o}s-R\'{e}nyi random graph (Fig.~\ref{plot_InfBd}f).

\begin{figure*} 
\includegraphics[width = 0.6\textwidth]{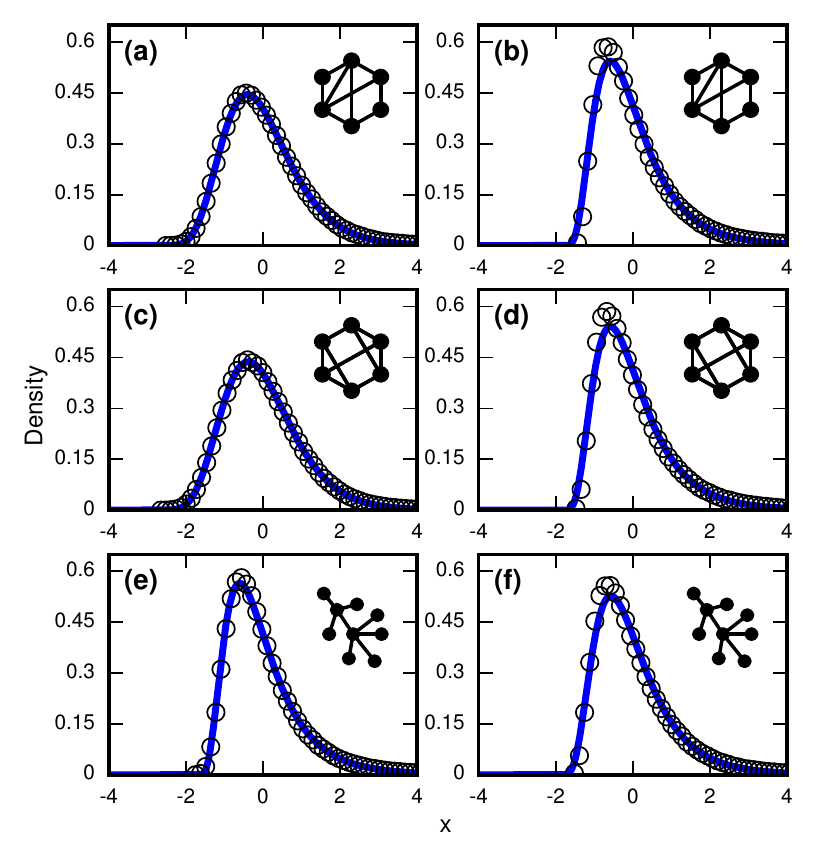} 
\caption{\textbf{Complex networks.} Simulated and fitted distributions of invader fixation times for Birth-death dynamics on small-world, scale-free, and $k$-regular networks. All means $\mu$ and variances $\sigma^2$ were computed empirically. The curves indicate non-central lognormals fitted to the first three moments of the data. All distributions are the result of $10^6$ simulations. The figures in the left column ((a), (c), (e)) used invader fitness $r = \infty$, whereas the figures in the right column ((b), (d), (f)) used neutral fitness $r = 1$. (a) Newman-Watts-Strogatz small-world ring network with shortcut probability of $\rho = 0.25$ on $N=75$. (b) Newman-Watts-Strogatz small-world ring network with shortcut probability of $\rho = 0.25$ on $N=25$. (c) Random 3-regular graph on $N=100$ nodes. (d) Random 3-regular graph on $N=22$ nodes. (e) Barabasi-Albert scale-free network with a minimum degree of 3 and $N=100$ nodes. (f) Barabasi-Albert scale-free network with a minimum degree of 3 and $N=22$ nodes. }
\label{plot_bonus_graphs}
\end{figure*}

For more complex networks, such as the Watts-Strogatz small-world network, the $k$-regular random graph, and the Barabasi-Albert scale-free network, we currently lack theory to predict the asymptotic distributions analytically. However, numerical simulations produce simulations which are all well-approximated by a noncentral lognormal, obeying Sartwell's law~\cite{sartwell1950distribution} (Fig.~\ref{plot_bonus_graphs}a,c,e).

Extended Data Table~\ref{table_Disp} shows that geometric standard deviations of the incubation period distributions for \emph{all} of these networks fall around $1.1-1.4$, in agreement with the  dispersion factors of $1.1-1.5$ observed for many infectious diseases~\cite{sartwell1950distribution, horner1992criteria}.   

\begin{table}
\begin{tabular}{|c||c|c|}
\hline
Fitness Level & $r = \infty$ & $r = 1$ \\
\hline
Complete                  &   $  1.2386  \pm 0.0004 $ & $  1.6629  \pm  0.0012 $  \\
\hline
Star                          &   $  1.3463  \pm 0.0006 $ & $  1.6875  \pm  0.0012 $  \\
\hline
1D Lattice                 &   $  1.1418  \pm 0.0002 $ & $  1.7907  \pm  0.0014 $  \\
\hline
2D Lattice                 &   $  1.0731 \pm 0.0003 $ & $   1.6799  \pm  0.0012 $  \\
\hline
3D Lattice                 &   $  1.1289 \pm 0.0006 $ & $   1.6659  \pm  0.0012 $  \\
\hline
Erd\H{o}s-R\'{e}nyi &   $  1.2586 \pm 0.0004 $ & $   1.6900  \pm   0.0012 $  \\
\hline
Small World              &   $  1.2604  \pm 0.0004 $ & $  1.7693  \pm  0.0014 $  \\
\hline
k-Regular                 &   $  1.2125  \pm 0.0003 $ & $  1.7229  \pm  0.0013 $  \\
\hline
Scale-Free               &   $  1.4189  \pm 0.0007 $ & $  1.7399  \pm  0.0013 $  \\
\hline
\end{tabular}
\caption{\textbf{Model dispersion factors.}  Dispersion factors (that is, geometric standard deviations) for the simulated distributions of incubation periods shown in Extended Data Figs.~\ref{plot_InfBd},~\ref{plot_bonus_graphs}, and~\ref{plot_Bd}, for different networks and invader fitness levels. Error bars represent 95\% confidence intervals. Due to finite size effects, the dispersion factors exceed 1 for 1D and 2D lattices with $r = \infty$ (they should approach 1 as $N \rightarrow \infty$).  Dispersion factors for the $r=1$ case are larger than for the $r = \infty$ case, but are more uniform for different network topologies.}
\label{table_Disp}
\end{table}

\subsection*{Random Walk Skewness}

So far we have focused on infinitely fit invaders ($r \rightarrow \infty$). Now we consider the opposite extreme, where invaders have nearly neutral fitness ($r \approx 1$) relative to the residents. We will show that right-skewed distributions of incubation periods occur in this limit as well, but for a completely different reason than coupon collection.

The analysis is again simplest for the complete graph, so we return to that case. As before, the probability of an invader replacing a resident in the next time step is 

\begin{equation*}
p_{m}^{+} := \frac{mr}{mr + (N-m)} \cdot \frac{N-m}{N-1} .
\end{equation*}

\noindent Similarly, the probability of an invader being replaced by a resident in the next time step is

\begin{equation*}
p_{m}^{-} := \frac{N-m}{mr + (N-m)} \cdot \frac{m}{N-1} .
\end{equation*}

\noindent So the probability of the next replacement adding a new invader is  

\begin{equation*}
q := \frac{p_{m}^{+} }{p_{m}^{+} + p_{m}^{-} } = \frac{r}{r+1}. 
\end{equation*}

\noindent This defines a random walk with drift $q$ on the invader population. 

\begin{figure*}[ht] 
\includegraphics[width = 0.6\textwidth]{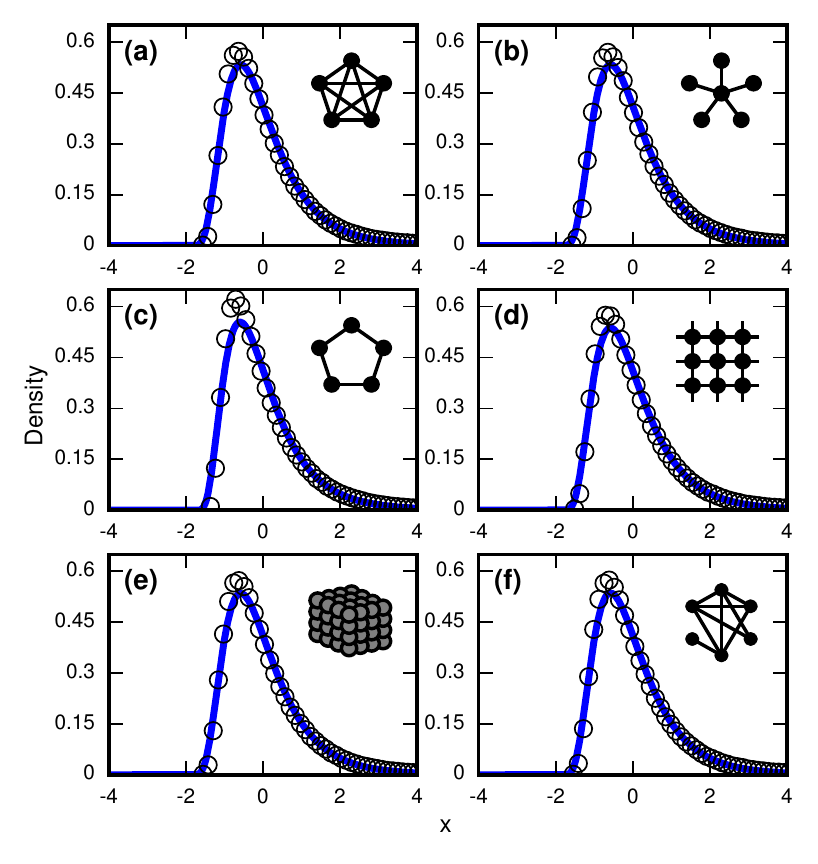} 
\caption{Neutrally fit invader ($r=1$). Simulated and fitted distributions of invader fixation times are shown for Birth-death dynamics on various networks. All means $\mu$ and variances $\sigma^2$ were computed empirically. The curves indicate noncentral lognormals fitted via the method of moments. (a) Complete graph on $N=50$ nodes, for $10^6$ runs. (b) Star graph with $N=25$ spokes, for $10^6$ runs. (c) One-dimensional ring on $N=50$ nodes, for $10^6$ runs. (d) Two-dimensional lattice on $N = 7 \times 7$ nodes, for $10^6$ runs. (e) Three-dimensional lattice on $N = 4^3$ nodes for $10^6$ runs. (f) Erd\H{o}s-R\'{e}nyi random graph on $N = 25$ nodes with an edge probability of $\rho = 0.5$. }
\label{plot_Bd}
\end{figure*}

But only a special subset of these walks are relevant to the computation of the incubation period distribution. For the incubation period to be well-defined, the invader population must not go extinct. Therefore, we need to condition on the fact that the invader population $m$ hits $N$ before it ever hits 0. For the limiting case $r=1$, corresponding to a perfectly neutral invader, we can show with martingale methods that the resulting distribution of incubation periods will be strongly skewed to the right as $N$ gets large (Supplementary Information). This is to be expected: there are only a few ways to walk from 1 to $N$ quickly, while there are many ways to have a long, meandering excursion before finally getting there. 

The variance from this conditioned random walk process tends to drown out the effects of network topology. The distribution of incubation periods ends up looking similar for diverse networks (Fig.~\ref{plot_Bd}), including complex networks  (Fig.~\ref{plot_bonus_graphs}b,d,f). So even though no coupon collection happens at low finesses $r \approx 1$, the effect of the conditioned random walk is more than enough to generate right-skewed distributions of incubation periods. In fact, this conditioned random walk mechanism at low $r$ produces an even higher dispersion factor ($\approx 1.7$) than coupon collection does at high $r$ (see Extended Data Table~\ref{table_Disp}). 

\subsection*{Testing robustness to update rule, fitness, and truncation}

\begin{figure}[ht] 
\includegraphics[width = 0.6\textwidth]{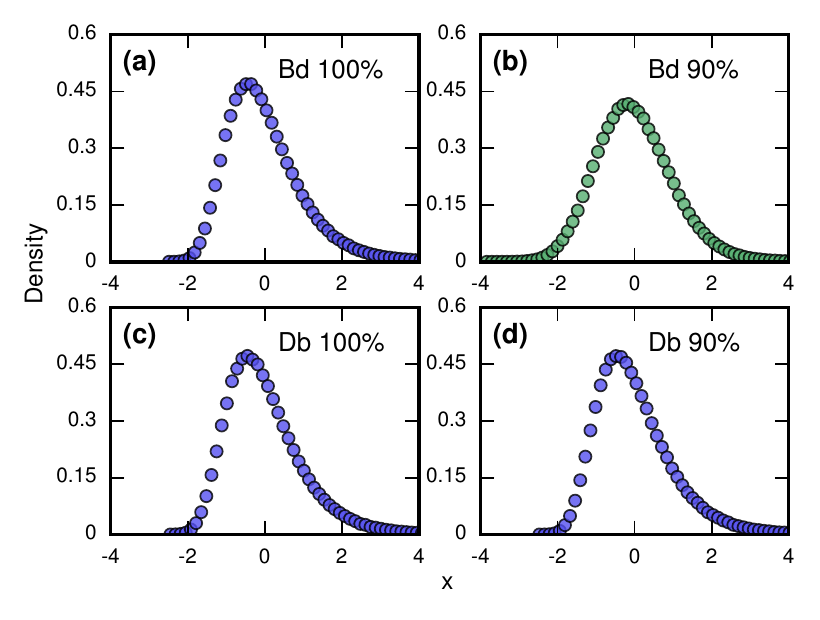} 
\caption{\textbf{Testing robustness.} The plots show how the distribution of incubation periods does or does not change when we modify the assumed update rules, invader fitness, and criterion for the onset of symptoms. Both Birth-death (Bd) and Death-birth (Db) dynamics were simulated on a complete graph of $N=500$ nodes, using an invader fitness of $r = 10$, intermediate between the limiting cases of infinite $r$ and $r=1$ considered earlier. Incubation periods are now defined as times needed for invaders to take over a fraction $f$ of the whole network. All distributions are normalized to have zero mean and unit variance. (a) The distribution of times till invader fixation ($f=1$) under Birth-death dynamics. The Gumbel distribution of Fig.~\ref{plot_MultiR}a persists for finite $r$. (b) When $f$ is reduced to 0.9, the incubation periods under Birth-death dynamics become normally distributed instead of skewed. By contrast, Death-birth dynamics are insensitive to this modification: the Gumbel distribution persists not only for (c) $f=1$ but even for (d) $f=0.9$. This insensitivity reflects when coupon collection occurs.  For Death-birth dynamics, it occurs near the \emph{beginning} of the invasion, when it takes a long time to randomly select one of the few available invaders to give birth. Since coupon collection occurs near the beginning, it is insensitive to the end-condition $f < 1$. In contrast, coupon collection occurs near the \emph{end} of the invasion for Birth-death dynamics (when residents are scarce), and hence gets truncated when $f < 1$, giving rise to a normal instead of a right-skewed distribution.}
\label{plot_MultiF}
\end{figure}

Right-skewed distributions typically persist in the face of various perturbations, but some perturbations can turn them into normal distributions. For example, suppose we allow intermediate invader fitness $1<r<\infty$ and allow symptoms to occur when invaders take over only a fraction $f$ of the whole network. (After all, leukemic cells need not take over all the bone marrow before leukemia becomes evident,
nor does typhoid need to overwhelm all the cells in the microbiome before causing fever.) Figure~\ref{plot_MultiF} contrasts what happens for Birth-death and Death-birth dynamics when we allow partial takeovers to trigger symptoms. When $r=10$, the Gumbel distribution of Fig.~\ref{plot_MultiR}a persists for $f=1$ (Fig.~\ref{plot_MultiF}a), but turns into a normal distribution~\cite{baum65} when $f=0.9$ (Fig.~\ref{plot_MultiF}b). Yet under Death-birth dynamics, the distribution stays Gumbel for both values of $f$ (Figs.~\ref{plot_MultiF}c, d). The fact that similar update rules behave so differently under a reasonable perturbation should caution us to choose our models with care. 

\subsection*{Influence of heterogeneity}

Historically, the distribution of incubation periods has been ascribed to heterogeneity~\cite{sartwell1950distribution, nishiura07,horner1992criteria} in the strength and dose of the pathogen, and in the immune response of the host. To see how these potential sources of heterogeneity could account for the skewed and approximately lognormal distribution of incubation periods, consider a pathogen growing exponentially with rate $r$ from an initial population $N_0$, so that its population at time $t$ is given by $N(t) = N_0 e^{r t}$.
If an immune response or other detectable symptoms are triggered when $N$ reaches a threshold population $\theta$, then the incubation time $T$ satisfies $N(T) = N_0 e^{r T} = \theta$. Solving for $T$ yields

\begin{equation}
T = \frac{1}{r} \left(\log\theta - \log N_0\right).
\end{equation}

\noindent So if either the threshold $\theta$ or the inoculum $N_0$ are normally distributed across the host population, the incubation period $T$ will be lognormally distributed.  Likewise, but in a more qualitative sense, a normal distribution of pathogen growth rates $r$ will also produce a skewed distribution that resembles a lognormal~\cite{nishiura07}.  However, if there is no randomness in any of those sources, this model predicts a single deterministic value of $T$ for the incubation period.  

In contrast, the stochastic model proposed here (see main text) does not need these sources of heterogeneity to produce right-skewed distributions. But if they happen to be present, as they likely are for many real outbreaks of infectious disease, our model can accommodate them. Indeed, when any of the three sources of heterogeneity are included in our model, they only serve to make the predicted distributions even more right-skewed, as we now show.

\begin{figure*} 
\includegraphics[width = 0.7\textwidth]{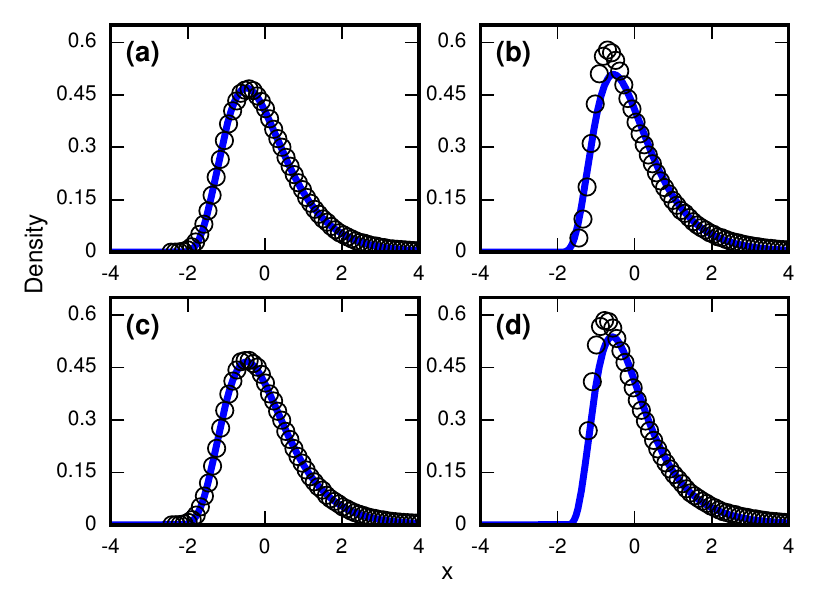} 
\caption{\textbf{Robustness to heterogeneity.} Simulated, fitted, and normalized distributions of incubation periods for Birth-death dynamics on a complete graph of $N=500$ nodes. Unless stated otherwise, each simulation used an invader fitness of $r=10$, measured times till complete takeover ($f=1$), and started from an initial dose of 1 invader. Runs where the dosage was not  smaller than the truncation point were rejected. The blue curves indicate noncentral lognormals fitted via the method of moments. (a) Heterogeneous fitness of invader. Every run used a different $r$ selected from a Gamma distribution with a shape parameter of 10. (b) Heterogeneity of host response. Instead of waiting until all $N$ residents had been replaced by invaders, every run used a different truncation point uniformly selected from $\{2, 3, \ldots N\}$. (c) Heterogeneity of dosage. Every run had a different starting population drawn from a Poisson of mean 10 and a shift of 1. (d) Heterogeneity of all three. Every run used an $r$ drawn from Gamma(10), a truncation point $f$ drawn from Uniform(0,1), and a dosage drawn from Poisson(10)+1.}
\label{plot_truncs_and_fits}
\end{figure*}

First, to emulate the heterogeneity of the strength of the pathogen, we assume heterogeneity in the parameter $r$ (which, in our model, governs the fitness of the invading cells relative to those of the host). In particular we randomly draw a different $r>0$ in each simulation, to simulate different hosts being infected with different pathogenic strains. The resulting distribution of invader fixation times depends on the distribution of the $r$'s, but our investigations demonstrate they consistently produce right-skewed distributions (Fig.~\ref{plot_truncs_and_fits}a).

Second, to emulate the heterogeneity of host immune responses, we allow variability in the parameter $f$, which quantifies the fraction of the network that needs to be invaded before symptoms appear. Let $T_f$ denote the time it takes for $N\cdot f$ of the original resident nodes to be replaced by invaders. If we draw $f$ randomly from some distribution, then essentially each host has a different threshold at which symptoms appear. In contrast to Fig.~\ref{plot_MultiF}b, where we saw that repeated simulations for a host population with a single, fixed, deterministic $f$ can cause skewed distributions to turn into normal distributions, that is no longer the case when heterogeneity is included, as Fig.~\ref{plot_truncs_and_fits}b indicates. In fact, the heterogeneity actually causes even more right-skew than before. 

Third, emulating variable doses is also straightforward. Instead of always starting with a single invader cell, we choose the initial number of invaders according to some distribution. Again, this modification does not remove the right-skewed behavior established in the Moran model (Fig.~\ref{plot_truncs_and_fits}c).

Finally, we can apply all these sources of heterogeneity at once, and remain with a right-skewed distribution (Fig.~\ref{plot_truncs_and_fits}d). 

In summary, although our results in the main text were based obtained by analyzing stochastic models of homogeneous host and pathogen populations, allowing for heterogeneity makes the predicted right-skewed distributions more, not less, prominent.

\appendix

\section{Agreement of geometric and exponential variables I} \label{Prop_GeoExpI}
\noindent {\bf Proposition: } {
Suppose we have a family of sequences $(p_m)_{m=1}^M$, with $0 \leq p_m \leq 1$ for all $m$ and $M$, where $p_m$ may depend on $M$. Define $\mbox{Geo}(p)$ to be a geometric random variable with distribution 
\begin{equation*}
P(\mbox{Geo}(p) = k) = (1-p)^{k-1} p
\end{equation*}
for $k = 1, 2, \ldots$. Further, let $\mathcal{E}(p)$ be an exponential random variable with distribution 
\begin{equation*}
P(\mathcal{E}(p) = x)dx = p e^{-p x} dx
\end{equation*}
for $x \geq 0$. Given some function $L := L(M)$ such that 
$\lim_{M\to \infty} L = \infty$
and 
\begin{equation}\label{GeoToExpConditionI}
\lim_{M \to \infty} \sum_{m=1}^M \frac{1}{p_mL^2} = 0,
\end{equation}
and given $T_G := \sum_{m=1}^M X(p_m)$, $T_E := \sum_{m=1}^M \mathcal{E}(p_m)$, and $\mu := \sum_{m=1}^M 1/p_m$, then
\begin{equation} \label{GeoToExpI}
\frac{T_G - \mu}{L} \sim \frac{T_E - \mu}{L} .
\end{equation}
The symbol ``$\sim$'' means the ratio of characteristic functions goes to 1 as $N$ gets large. That is, the random variables on both sides converge to each other in distribution as $M$ gets large.
}

\noindent {\bf Proof:} The proof of this claim simply involves calculating the characteristic functions and taking a limit.  We have presented the details elsewhere~\cite{ottinoloffler2017takeover}.

\section{Agreement of geometric and exponential variables II} \label{Prop_GeoExpII}
\noindent {\bf Proposition: } {
Given the setup in the previous proposition, define $\sigma_{G}^2 = \mbox{Var}(T_G)$ and $\sigma_{E}^2 = \mbox{Var}(T_E)$. If
\begin{equation} \label{GeoToExpConditionII}
\lim_{M \to \infty} \frac{ \sum_{m=1}^M p_{m}^{-1} }{ \sum_{m=1}^M p_{m}^{-2} } = 0,
\end{equation}
then 
\begin{equation} \label{GeoToExpII}
\frac{T_G - \mu}{\sigma_G} \sim \frac{T_E - \mu}{\sigma_E} .
\end{equation}

\noindent {\bf Proof:} Our first goal is to show that 
\begin{equation*} 
\frac{T_G - \mu}{\sigma_G} \sim \frac{T_E - \mu}{\sigma_G}.
\end{equation*}
We do this by using proposition~\ref{Prop_GeoExpI}, substituting $\sigma_G$ for $L$. First we check that Eq.~\eqref{GeoToExpConditionI} is satisfied. Notice that
\begin{alignat*}{1}
\lim_{M \to \infty} \sum_{m=1}^M \frac{1}{p_m\sigma_{G}^2} &= \lim_{M \to \infty} \frac{\sum_{m=1}^M p_{m}^{-1} }{ \sum_{m=1}^M p_{m}^{-2} - p_{m}^{-1} } \\
&= \lim_{M \to \infty} \frac{ \sum_{m=1}^M p_{m}^{-1} / \sum_{m=1}^M p_{m}^{-2} }{1 - \sum_{m=1}^M p_{m}^{-1} / \sum_{m=1}^M p_{m}^{-2} } \\ 
&= \frac{0}{1-0} = 0
\end{alignat*}
by hypothesis. Hence 
\begin{equation*} 
\frac{T_G - \mu}{\sigma_G} \sim \frac{T_E - \mu}{\sigma_G} = \frac{\sigma_E}{\sigma_G} \frac{T_E - \mu}{\sigma_E}.
\end{equation*}
But notice that
\begin{alignat*}{1}
\lim_{M \to \infty} \frac{\sigma_{E}^2}{\sigma_{G}^2} &= \lim_{M \to \infty} \frac{\sum_{m=1}^M p_{m}^{-2} }{ \sum_{m=1}^M p_{m}^{-2} - p_{m}^{-1} } \\
&= \lim_{M \to \infty} \frac{ 1 }{1 - \sum_{m=1}^M p_{m}^{-1} / \sum_{m=1}^M p_{m}^{-2} } \\ 
&= 1.
\end{alignat*}
Therefore, the proposition is proven.

\section{Condition for normality} \label{GeneralNormal}
\noindent {\bf Proposition: } {
Let $T = \sum_{m=1}^{M} \mathcal{E}(p_m)$, define $\sigma^2 = \mbox{Var}(T) = \sum_{m}^{M} p_{m}^{-2},$ and let $\lim_{M \to \infty} p_{m} \sigma = \infty$. If
\begin{equation} \label{ExpToNormalpCondition}
\lim_{M \to \infty} \sum_{m=1}^M \exp\left(-\epsilon p_m \sigma \right) = 0,
\end{equation}
then 
\begin{equation} \label{ExpToNormal}
\frac{T - \mu}{\sigma} \xrightarrow{d} \mbox{Normal}(0,1).
\end{equation}

\noindent {\bf Proof:} Apply the Lindeberg-Feller central limit theorem~\cite{durrett91} to the random variables 
\begin{equation*}
Y_{m,M} := \frac{\mathcal{E}(p_m) - 1/p_m}{\sigma}.
\end{equation*}
By construction, $\sum_m(Y_m,M) = (T-\mu)/\sigma$, $E[Y_{m,M}] = 0$ and $\sum_m E[Y_{m,M}^2] = 1$. So in order to apply the theorem (and thereby get our desired result), we simply need satisfy the Lindeberg condition for any $\epsilon > 0$, as given by 
\begin{equation*}
\lim_{M \to \infty} \mbox{Lind}_M := \lim_{M \to \infty} \sum_{m=1}^M E[Y_{m,M}^2; |Y_{m,M}| > \epsilon] = 0.
\end{equation*}
Notice that $Y_{m,M} < -\epsilon$ implies 
\begin{alignat*}{1}
\E(p_m) < p_{m}^{-1} - \epsilon \sigma_{E}^2 = p_{m}^{-1}( 1- \epsilon p_m \sigma).
\end{alignat*}
By hypothesis, the right hand side will eventually be less than 0, meaning that eventually $Y_{m,M} < -\epsilon$ will be impossible. So if we define $c_m = 1 + \epsilon p_m \sigma$, then we know that 
\begin{alignat*}{1}
\lim_{M \to \infty} \mbox{Lind}_M = \lim_{M \to \infty} \sum_{m=1}^M E[Y_{m,M}^2; Y_{m,M} > \epsilon].
\end{alignat*}
So for large enough $M$, we have 
\begin{alignat*}{1}
& \mbox{Lind}_M \\
& := \sum_{m=1}^M \int_{c_m/p_m}^\infty \left( \frac{x - 1/p_m}{\sigma} \right)^2 e^{-p_m x} p_m dx \\
& = \sum_{m=1}^M \frac{1}{\sigma^2 p_{m}^2} \int_{c_m}^\infty \left(y-1\right)^2 e^{-y} dy \\ 
& = \sum_{m=1}^M \frac{1}{\sigma^2 p_{m}^2} e^{-c_m} (c_{m}^2 +1) \\
& = \sum_{m=1}^M \frac{1}{e \sigma^2 p_{m}^2} e^{-\epsilon p_m \sigma} (2 + 2\epsilon p_m \sigma + \epsilon^2 p_{m}^2\sigma^2).
\end{alignat*}
By hypothesis, $p_m \sigma$ grows without bound, so the term  $p_{m}^2\sigma^2$ will be dominant. Therefore, there is some constant $D$ such that we can create the upper bound
\begin{alignat*}{1}
\mbox{Lind}_M &\leq \sum_{m=1}^M \frac{1}{\sigma^2 p_{m}^2} e^{-\epsilon p_m \sigma} D \sigma^2 p_{m}^2 \\
& \leq D \sum_{m=1}^M \exp\left(-\epsilon p_m \sigma \right).
\end{alignat*}
From here, we can apply the hypothesis to get 
\begin{alignat*}{1}
\lim_{M \to \infty} \mbox{Lind}_M \leq \lim_{M \to \infty} D \sum_{m=1}^M \exp\left(-\epsilon p_m \sigma \right) = 0.
\end{alignat*}
Thus the Lindeberg condition holds. Therefore, by applying the theorem, we conclude that 
\begin{equation*}
\sum_m(Y_{m,M}) = \frac{T-\mu}{\sigma} \xrightarrow{d} \mbox{Normal}(0,1).
\end{equation*}

\section{Normally distributed fixation times for 1D lattice} \label{1D_Distribution}

Let us start with a one-dimensional (1D) ring of $N$ nodes, with exactly one invader  at the start. Under infinite-$r$ Birth-death (Bd) dynamics, a uniform random invader gives birth at every time step and replaces one of its neighbors, also uniformly at random. 

Because of the simple topology of the ring, the growing chain of invaders will always advance from the left or right ends. This means that if there are currently $m$ invaders, then the probability of a new invader being added in this time step is exactly given by
\begin{equation*}
p_m = \frac{1}{m}
\end{equation*}
for $m = 1, 2, \ldots, M$, where we have defined $M := N-1$. 

The probability of spending exactly $t$ time steps at $m$ invaders is given by the odds of doing nothing for exactly $t-1$ steps and then advancing at the last step, and so is given by $(1-p_m)^{t-1}p_m$. In other words, the time spent with $m$ invaders is given by the geometric random variable $\mbox{Geo}(p_m)$. Therefore the  total fixation time $T$ is given by $T = \sum_{m=1}^M \G(p_m)$. 

By applying the results of section~\ref{Prop_GeoExpII} above, we can switch to using exponential random variables. From here we wish to apply condition~\ref{GeneralNormal}, so we need to check if $\sum_{m=1}^M\exp(-\epsilon p_m \sigma)$ converges to zero, given $\sigma = \sum_{m=1}^M p_{m}^{-2}$. Using asymptotics and a constant $D$, we find
\begin{alignat*}{1}
\sum_{m=1}^M\exp(-\epsilon p_m \sigma) &\leq \sum_{m=1}^M \exp\left(\frac{-D}{m}M^{3/2} \right) \\ 
& \leq \sum_{m=1}^M \exp\left(\frac{-D}{M}M^{3/2} \right) \\
& = M \exp\left( -D \sqrt{M}\right) \to 0. 
\end{alignat*} 
This lets us cite our proposition, and conclude that the fixation time is asymptotically distributed as a normal, with 

\begin{equation*}
\sum_m(Y_{m,M}) = \frac{T-\mu}{\sigma} \xrightarrow{d} \mbox{Normal}(0,1).
\end{equation*}

\section{Gumbel distributed fixation times for star graph} \label{Star_Distribution}

Next consider the infinite-$r$ Bd dynamics on a star graph. This network consists of one ``hub'' node and $N$ ``spoke'' nodes, with edges exclusively between the hub and spokes. We will place the initial invader at the hub, since starting from a spoke is a trivial perturbation off of that. 

Now, given that $m$ of the $N$ spokes are invaders, then the odds of another one turning into an invader in one time step is simply the odds of choosing the hub from the set of all invaders, times the probability of replacing an existing healthy spoke. So
\begin{equation*}
p_m = \frac{1}{m+1} \cdot \frac{N-m}{N}
\end{equation*}
for $m = 0, 1, \ldots, N-1$. As before, the fixation time $T$ is the sum of geometric random variables $\G(p_m)$. However, it is easy to use section \ref{Prop_GeoExpI} to show that the total fixation time is well approximated by the sum of exponential random variables $\E(p_m),$ given we normalize by $N^2$. So we know 
\begin{equation} \label{Star_GeoToExp}
\frac{T - E(T)}{N^2} = \sum_{m=1}^{N-1} \frac{\G(p_m) - 1/p_m}{N^2} \sim \sum_{m=1}^{N-1} \frac{\E(p_m) - 1/p_m}{N^2}. 
\end{equation}

The crux of finding the limiting distribution of $T$ is noticing that $N p_m \approx (N-m)/N$ for large $m$. The sequence $r_m := (N-m)/N, m = 0, \ldots, N-1$ corresponds to the well-known {\it coupon collector's problem} (see main text). Imagine a child trying to complete a collection of $N$ cards by buying one random card each week. The probability of getting a new card the first week is 1; the probability of getting a new card after the first has been collected is $(N-1)/N$; the probability of getting another new card after two cards have been collected is $(N-2)/N$; and so on until the probability of getting the last card is $1/N$.

The time $T_C$ to complete this collection (which is also well approximated by the sum of exponentials) has been the subject of much historical study. In fact, an exact distribution for $T_C$ in the limit of large $N$ is known~\cite{ottinoloffler2017takeover, posfai10,feller1968introduction, erdos61, baum65, rubin65}, and is given in normalized form by 
\begin{equation}\label{Coupon_Limit}
\frac{T_C - E(T_C)}{N} \sim \sum_{m=1}^{N-1} \frac{\E(r_m) - 1/r_m}{N} \xrightarrow{d} \mathrm{Gumbel}(-\gamma, 1),
\end{equation}
where $\gamma \approx 0.5772$ is the Euler-Mascheroni constant.

Therefore, if we can connect our fixation time $T$ to the coupon collector's time $T_C$, we would know its distribution. We do so by taking the ratio of their respective characteristic functions, using their respective approximations as exponential random variables. Letting $k = N-m$, we find a characteristic function 
\begin{alignat*}{1}
& \phi := \\
& = E\left[ \exp\left( it \sum_{k=1}^N \frac{\E(p_k) -1/p_k}{N^2} \right) \right] \\
& =\prod_{k=1}^N \frac{\exp{-it/(N^2p_k)}}{1 - it/(N^2p_k)} \\
\end{alignat*}
for our fixation time, and 
\begin{alignat*}{1}
& \phi_C := \\
& = E\left[ \exp\left( it \sum_{k=1}^N \frac{\E(r_k) -1/r_k}{N} \right) \right] \\
& =\prod_{k=1}^N \frac{\exp{-it/k}}{1 - it/k} \\
\end{alignat*}
for the coupon collector's time. 

Taking the ratio gives 
\begin{alignat*}{1}
\frac{\phi_C}{\phi} & = \prod_{k=1}^N \exp\left[ \frac{-it}{N}\left(1 - \frac{1}{k} \right) + \log\left( 1+ \frac{it}{N} \frac{k-1}{k-it} \right) \right].
\end{alignat*}

Taking the Taylor expansion of the logarithm for $N \gg 1$, we can substitute in an appropriate function $R_m$ which gets small as $N$ gets large. Thus
\begin{alignat*}{1}
\frac{\phi_C}{\phi} & = \exp\left[ \frac{it}{N} \sum_{k=1}^N \frac{it(k-1)}{k(k-it)}+ \frac{t^2}{2} \sum_{k=1}^N \frac{R_m}{N^2} \left( \frac{k-1}{k-it} \right)^2 \right]. 
\end{alignat*}

The first sum is bounded above in norm by a constant times $\log(N)$, so the first term goes to zero as $N$ gets large. Similarly, the second term goes to zero quickly, meaning that $\phi_C/\phi \to 0$. Hence 
\begin{equation} \label{Star_Crux}
\sum_{k=1}^N \frac{\E(p_k) -1/p_k}{N^2} \sim \sum_{k=1}^N \frac{\E(r_k) -1/r_k}{N}.
\end{equation}
Using Eq.~\eqref{Star_Crux} to connect Eq.~\eqref{Star_GeoToExp} to Eq.~\eqref{Coupon_Limit}, we find
\begin{equation}\label{Star_Limit}
\frac{T - E(T)}{N^2} \xrightarrow{d} \mathrm{Gumbel}(-\gamma, 1),
\end{equation}
as desired.

\section{Normally distributed fixation times for 2D lattice} \label{2D_Distribution}

We wish to find the limiting distribution of fixation times of infinite-$r$ Bd growth on a $d$-dimensional square lattice, assuming periodic boundaries. We will eventually focus on the two-dimensional (2D) lattice, but let us set up every case for $2 \leq d < \infty$ right now. 

Unlike the previous cases, the probability of adding a new invader always depends on the exact configuration of the existing invaders. So we can no longer define exact values for $p_m$ that describe the fixation time $T$ as a simple sum of random variables. 

But even though we do not know the exact shape of the invader cluster, the simple network structure can motivate a reasonable approximation. In particular, given a sufficiently smooth and convex volume $V$ in a $d$-dimensional lattice, we should expect the volume to have a surface area proportional to $V^{\eta}$, where $\eta = 1 - 1/d$. 

Assuming this to be true, recall the basic dynamics of infinite-$r$ Bd with $N$ nodes and $m$ invaders. First, we uniformly select one node out of the population of invaders, which is always a probability of 1/$m$ per node. Then we replace one of the invader's neighbors, uniformly at random. However, only invaders on the surface of the cluster even have a chance at replacing a healthy node!

Given sufficient regularity of the boundary of the cluster, this means that the probability of a invader replacing a healthy node is proportional to 
\begin{equation*}
q_m = \frac{1}{m} \cdot \mbox{Surface area of the invader cluster}. 
\end{equation*}

Using this logic, we expect the probability of adding an invader to go as $m^\eta/m$ at the start. However, remember we have a periodic lattice, so using $m^\eta$ as the surface area of the cluster stops being true halfway through, and using the smaller healthy cluster's volume is a better approximation. In other words, the surface area grows as $m^\eta$ at the start when the cluster begins forming, and shrinks as $(N-m)^\eta$ at the end when there are only a few healthy cells left.

This intuitive reasoning suggests that the ``true'' probability of adding a new invader, given that there are already $m$ of them, should be roughly proportional to
\begin{equation}\label{dDLattice_pm}
q_m = \frac{\min(m, N-m)^\eta}{m}.
\end{equation}
The fact that we only estimated $q_m$ up to proportionality is adequate, because when we use section~\ref{Prop_GeoExpII}, any multiplicative factors will just be absorbed by the variance anyway when we write down the normalized distribution. 

The case of $d = 2$ is special, so let us plug in $d=2$ and $\eta = 1/2$ where appropriate. This gives 
\begin{equation*}
q_m = \frac{\min(m, N-m)^{1/2}}{m}.
\end{equation*}
Now use section~\ref{Prop_GeoExpII}. Skip ahead to defining $T$ to be the sum of exponential random variables, and split the sum in half. Thus
\begin{equation*}
T := T_a + T_b := \sum_{m=1}^{N/2-1} \E(q_m) + \sum_{m = N/2}^{N-1} \E(q_m).
\end{equation*}
We assume $N$ is even without loss of generality.

First, we show that $T_a$ is normally distributed using section~\ref{GeneralNormal}. To use this result we first need to calculate $\mbox{Var}(T_a)$. This is exactly
\begin{alignat*}{1}
\mbox{Var}(T_a) = \sum_{m=1}^{N/2-1} q_{m}^{-2} = \sum_{m=1}^{N/2-1} m = \frac{(N/2-1)^2 + (N/2-1)}{2}.
\end{alignat*}
Therefore, we have
\begin{alignat*}{1}
q_{m}^2 \mbox{Var}(T_a) = \frac{1}{m} \frac{(N/2-1)^2 + (N/2-1)}{2} \geq \frac{1}{m} \frac{N^2}{8} \geq \frac{1}{N} \frac{N^2}{8} = \frac{N}{8} \to \infty
\end{alignat*}
as $N \rightarrow \infty$, so the first condition is satisfied. 

Next, we need to show that a certain sum of exponentials converges to zero. In particular, for any $\epsilon > 0$, we examine
\begin{alignat*}{1}
S_N := \sum_{m=1}^{N/2 - 1} \exp\left(-\epsilon q_m \sqrt{\mbox{Var}(T_a)}\right).
\end{alignat*}
We can make the bound $\sqrt{\mbox{Var}(T_a)} > N/8$ and $q_m = 1/\sqrt{m} > 1/\sqrt{N}$, so 
\begin{alignat*}{1}
S_N \leq \sum_{m=1}^{N/2 - 1} \exp\left(-\epsilon \sqrt{N}/8 \right) \leq N \exp\left(-\epsilon \sqrt{N}/8 \right) .
\end{alignat*}
Therefore $S_N \to 0$ as $N$ gets large. So the second condition is satisfied. Therefore, $T_a$ is distributed according to a normal. 

However, we need to do the same to $T_b$. So let us estimate the variance of this second contribution. Letting $k = N-m,$ we get 
\begin{alignat*}{1}
\mbox{Var}(T_b) &= \sum_{k=1}^{N/2} q_{N-k}^{-2} = \sum_{k=1}^{N/2} \left( \frac{N-m}{\sqrt{k}} \right)^{2} \\ 
&= N^2\left( \sum_{k=1}^{N/2} \frac{1}{k} -\frac{2}{N} \sum_{k=1}^{N/2} 1 + \frac{1}{N^2} \sum_{k=1}^{N/2} k \right) \\
&= N^2\left( \sum_{k=1}^{N/2} \frac{1}{k} - 1 + \frac{1}{8} + \frac{1}{8N}\right).
\end{alignat*}
So for large $N$, we obtain the following bound: 
\begin{equation*}
\mbox{Var}(T_b) \geq \frac{N^2}{4} \log(N).
\end{equation*}
Therefore,  
\begin{alignat*}{1}
q_{k}^2 \mbox{Var}(T_b) \geq \frac{k}{(N-k)^2} \frac{N^2}{4} \log(N) \geq \frac{1}{4} \log{N}.
\end{alignat*}
This satisfies the first condition in section~\ref{GeneralNormal}. 

To satisfy the second condition, we again fix some arbitrary $\epsilon > 0$ and calculate a certain sum of exponentials. By calculating and choosing careful bounds, we get
\begin{alignat*}{1}
S_N & := \sum_{k=1}^{N/2} \exp\left(-\epsilon q_k \sqrt{\mbox{Var}(T_b)} \right) \\ 
& \leq \sum_{k=1}^{N}\exp\left(-\frac{\epsilon}{2} \frac{\sqrt{k}}{N-k} N \sqrt{\log(N)} \right) \\
& \leq \sum_{k=1}^{N}\exp\left(-\frac{\epsilon}{2} \sqrt{k \log(N)} \right)
\end{alignat*}
This sum can be approximated from above by an appropriate integral:
\begin{alignat*}{1}
S_N & \leq \int_{0}^\infty\exp\left(-\frac{\epsilon}{2}\sqrt{x \log(N)}\right) dx \\ 
& = \frac{8}{\epsilon^2 \log(N)}. 
\end{alignat*}
Therefore, $S_N \to 0$ as $N$ gets large. This satisfies the second condition in section~\ref{GeneralNormal}, meaning that we now know that $T_b$ is distributed as a normal.

Since the sum of normal variables returns a normal variable, this means that $T=T_a+T_b$ is also normal. Hence, we expect for the fixation time of infinite-$r$ Bd on a 2D lattice to be distributed as a normal, like the 1D ring lattice but, as we will now show, unlike $d \geq 3$.

\section{Non-normality for $d \geq 3$} \label{Lattice_Distribution}

Here, we wish to find the limiting distribution of fixation times of infinite-$r$ Bd growth on a $d$-dimensional square lattice, assuming periodic boundaries. Right now, we will look only at $3 \leq d < \infty$, since $d = 1, 2,$ and $\infty$ are special cases. 

We did the bulk of the setup for this case in section~\ref{2D_Distribution}, so we have the approximate probabilities of adding an invader to be given by 
\begin{equation*}\label{dDLattice_pm}
q_m = \frac{\min(m, N-m)^\eta}{m}
\end{equation*}
as before. And again, we define the ``approximate'' fixation time $T$ to be the sum of the exponential random variables $\E(q_m)$. Splitting the sum into a front and back half gives 
\begin{equation*}
T := T_a + T_b := \sum_{m=1}^{N/2-1} \E(q_m) + \sum_{m = N/2}^{N-1} \E(q_m).
\end{equation*}

But even with such aggressive approximations, we cannot present a closed form for the distribution of $T$. However, we can still calculate an important quantity: the skew of the distribution. 

Since we use section~\ref{Prop_GeoExpII}, let us skip to defining $T$ to be the sum of exponential random variables, and split the sum in half, as we did with the 2D lattice. Thus
\begin{equation*}
T := T_a + T_b := \sum_{m=1}^{N/2-1} \E(q_m) + \sum_{m = N/2}^{N-1} \E(q_m).
\end{equation*}
We assume $N$ is even without loss of generality.

First, let us find which half contributes more variance. Bound Var($T_a$) as
\begin{alignat*}{1}
\mbox{Var}(T_a) &= \sum_{m=1}^{N/2-1} q_{m}^{-2} = \sum_{m=1}^{N/2-1} m^{2/d} \\
& \leq \int_{0}^{N/2} x^{2/d} dx \leq N^{2/d + 1}.
\end{alignat*}
We similarly approximate Var($T_b)$, setting $k = N-m$ and finding 
\begin{alignat*}{1}
\mbox{Var}(T_b) &= \sum_{k=1}^{N/2} q_{N-k}^{-2} = \sum_{k=1}^{N/2} \frac{(N-k)^{2}}{k^{2\eta}} \\
& =N^2 \left( \sum_{k=1}^{N/2} \frac{1}{k^{2\eta}} -\frac{2}{N} \sum_{k=1}^{N/2} k^{1-2\eta}+ \frac{1}{N^2} \sum_{k=1}^{N/2} k^{2/d} \right) .
\end{alignat*}
Since $\eta \geq 2/3$, only the first term survives as $N$ gets large, so 
\begin{equation} \label{Var_b}
\mbox{Var}(T_b) \to N^2 \zeta(2\eta)
\end{equation}
where $\zeta$ is the usual Riemann zeta function. Notice that $2 > 2/d +1$ for $d \geq 3$. 

To use both these variances, recall the skewness summation formulation: if we have random variables $X_i$ with variances $\sigma^{2}_i$ and skews $\kappa_i$, then their sum has a skewness of 
\begin{equation} \label{SkewSum}
\mbox{Skew}\left(\sum_i X_i\right) = \frac{\sum_i \kappa_i \sigma_{i}^3}{\left(\sum_i \sigma_{i}^2\right)^{3/2}}.
\end{equation}
So this means that 
\begin{alignat*}{1}
\mbox{Skew}(T) & = \frac{\mbox{Skew}(T_a) \mbox{Var}(T_a)^{3/2} + \mbox{Skew}(T_b) \mbox{Var}(T_b)^{3/2} }{\left( \mbox{Var}(T_a) + \mbox{Var}(T_b) \right)^{3/2} } \\
& = \frac{\mbox{Skew}(T_a) (\mbox{Var}(T_a)/\mbox{Var}(T_b))^{3/2} + \mbox{Skew}(T_b) }{\left( 1+ \mbox{Var}(T_a)/\mbox{Var}(T_b) \right)^{3/2} } \\
& \to \mbox{Skew}(T_b)
\end{alignat*}
as $N \rightarrow \infty$. Here we have used the fact that $T_a$ has a finite skew (actually, it is easy to use section~\ref{GeneralNormal} to show that $T_a$ is distributed as a normal, and thus has zero skew.) Hence the asymptotic skew of $T$ is just the asymptotic skew of $T_b$. 

We can calculate the skew of $T_b$ by reusing Eq.~\eqref{SkewSum}, this time on the exponential variables defining $T_b$. Therefore
\begin{alignat*}{1}
\mbox{Skew}(T_b) = \frac{ \sum_{k=1}^{N/2} 2 q_{N-k}^{-3} }{\left( \sum_{k=1}^{N/2} q_{N-k}^{-2} \right)^{3/2}}. 
\end{alignat*}
By Eq.~\eqref{Var_b}, the denominator limits to $N^3 \zeta(2\eta)^{3/2}$. Meanwhile, the numerator looks like
\begin{alignat*}{1}
& 2 \sum_{k=1}^{N/2}\frac{1}{k^{3\eta}} \left(N-k \right)^3 \\
&= 2N^3 \left( \sum_{k=1}^{N/2} \frac{1}{k^{3\eta}} - \frac{3}{N} \sum_{k=1}^{N/2} k^{1-3\eta} + \frac{3}{N^2} \sum_{k=1}^{N/2} k^{2-3\eta} + \frac{1}{N^3} \sum_{k=1}^{N/2} k^{3/d} \right).
\end{alignat*}
The first term in the parentheses converges to $\zeta(3\eta)$, whereas the rest of the terms converge to 0. Combining the numerator and denominator gives us the conclusion that the skew for the full distribution is given by
\begin{equation} \label{dDLattice_Skew}
\mbox{Skew}(T) = \frac{2 \zeta(3\eta)}{\zeta(2\eta)^{3/2}}.
\end{equation}

Recall that $\eta = 1 - 1/d$, so the denominator diverges for $d=2$ and both then numerator and denominator diverge for $d = 1$. However, since this expression for $\mbox{Skew}(T)$ is monotone increasing in $\eta$ for $2/3 \leq \eta < 1$, every dimension $d \geq 3$ will attain a unique skew, and therefore a unique limiting distribution. Moreover, if we take $d \to \infty$, then $\eta \to 1$, and therefore the skew becomes $12\sqrt{6}\zeta(3)/\pi^3$, which is exactly the skew for a Gumbel distribution, supporting our assertion that high-dimensional systems attain higher skews. 

For the purposes of estimating the distributions at finite $N$, it is convenient to use the random variable 
\begin{equation*}
F(d) := \frac{ \sum_{k=1}^N \E\left(p_k\right) - p_{k}^{-1} }{\sum_{k=1}^N p_{k}^{-2} }
\end{equation*}
where
\begin{equation*}
p_k = \frac{k^{1-1/d}}{N-k}.
\end{equation*}
Since the second half of the dynamics contribute the majority of the variance, we should expect this to provide a reasonable approximation of $T$.

\section{Asymptotic skew of conditioned random walk} \label{RandomWalk_Distribution}

When $r=1$ for Bd dynamics on the complete graph, the number of invaders becomes very flexible. In fact, the probability of adding an invader on any time step is exactly equal to the probability of removing an invader, so
\begin{equation*}
p^{+}_m = \frac{N-m}{N} = p^{-}_m.
\end{equation*}
Hence the probability that the next event increases the number of invaders is always 1/2. This means that, if we ignore the waiting times, the population of invaders obeys a simple random walk on the values $m = 1, ..., N-1$ with 0 and $N$ acting as absorbing states. So we can understand the times required to take over the network by understanding these simple dynamics. To set up the analysis, let $X_n$ denote the number of invaders at time $n$. Therefore
\begin{equation*}
X_n = \sum_{i=1}^n x_i
\end{equation*}
where $x_i \in \{-1, +1\}$, each with probability 1/2. 

Although we always start at $m=1$, we only care about the invader takeover result because that is the only case for which disease symptoms would be manifested. Let us define the stopping time
\begin{equation*}
S_m = \min\{n | X_n = m \},
\end{equation*}
which records the first time the random walk $X_n$ hits the value $m$. Given that the invader population cannot go negative or above $N$, the walk's stopping time is 
\begin{equation*}
S = \min(S_0, S_N).
\end{equation*}
In the main text, we cared only about the conditioned random walk times and whether they tended to be right skewed. So we now study  the first few conditional moments 
\begin{equation*}
\mu_i := E(S^i | X_S = N),
\end{equation*}
for $i = 1, 2, 3$. 

It isn't hard to set up a linear recurrence relation to find the probability of hitting 0 or $N$. In fact, if the state of the walk is at $m$, then the probability of hitting $N$ is exactly
\begin{equation*}
P(X_S = N|X_n = m) = m/N.
\end{equation*}
This is a useful fact for simulation; instead of directly simulating $X_n$ and discarding all the cases that hit 0, we can directly simulate the conditioned random walk. If we define
\begin{equation*}
Y_n = E(X_n | X_S = N),
\end{equation*}
and treat it as a Markov chain, then we can easily calculate the transition probabilities. By applying Bayes's law,
\begin{alignat*}{1}
P(Y_{n} = m \to Y_{n+1} = m+1) &= \frac{P(X_n = m \to X_{n+1} = m+1 \mbox{ AND } X_S = N) }{P(X_S = N)} = \frac{1}{2}\frac{P(X_S = N|X_n = m+1)}{P(X_S = N|X_n = m) } \\ 
&= \frac{1}{2} \frac{m+1}{m}. 
\end{alignat*}
While this speeds up simulations by a good deal in certain cases, it is not particularly useful for quantifying the distribution of the random walk times themselves.

To identify the moments of $T$, we will want to apply results from martingale theory in general, and optional stopping in particular. To start, define the random variable
\begin{equation*}
M_n^{(1)} := X_{n}^3 - 3nX_n.
\end{equation*}
We are going to want this to be a martingale. Let's define $\F_n$ to be the sigma field consisting of all information from the first $n$ steps of the random walk. Therefore, $E(x_{n+1} | \F_n) = 0$, since we cannot predict the direction of the next step. However, we do know that $E(x_{n+1}^2 | \F_n) = 1$, because the steps will always be size 1, regardless of our ignorance. Putting this together gives
\begin{alignat*}{1}
E(M_{n+1}^{(1)} |\F_n) &= E\left( X_{n+1}^3 - 3(n+1)X_{n+1} | \F_n \right) \\
& = E\left( (X_n + x_{n+1})^3 - 3(n+1)(X_n + x_{n+1}) \right) \\
&= E\left( X_{n}^3 + 3 x_{n+1}X_{n}^2 + 3 x_{n+1}^2 X_n - 3(n+1)X_n -3(n+1)x_{n+1} \right) \\
&=X_{n}^3 -3nX_n \\
&= M_{n}^{(1)}.
\end{alignat*}
So $A_n^{(1)}$ well-approximates its future, meaning that it is a proper martingale. 

Thank to this, we can cite an optional stopping theorem~\cite{durrett91}. So we expect the expectation of this variable to be the same at the stopping time as at the start, so 
\begin{equation}\label{M1_equality}
E\left(M_{0}^{(1)}\right) = E\left(M_{S}^{(1)}\right).
\end{equation}
Since we start at $n=0$ and $X_0 = 1$, the left hand side trivially gives 
\begin{equation*}
E\left(M_{0}^{(1)}\right) = 1^3 - 3\cdot0\cdot1 = 1.
\end{equation*}
However, the right hand side gives something a bit more complicated, since we need to condition on the possible endpoints, remembering we start at $X_0 = 1$. So
\begin{alignat*}{1}
E\left(M_{S}^{(1)}\right) &= P(X_S = N)E(M_{S}^{(1)} | X_S = N) + P(X_S = 0)E(M_{S}^{(1)} | X_S = 0) \\
& = \frac{1}{N} E\left( X_{n}^3 - 3nX_n | X_S = N\right) + \frac{N-1}{N} E\left( X_{n}^3 - 3nX_n | X_S = 0\right) \\
&= \frac{1}{N} \left( N^3 - 3\mu_1N \right) + \frac{N-1}{N} E\left( 0^3 - 3E(S|X_S = 0)0 \right) \\
&= N^2 -3 \mu_1.
\end{alignat*}

Thus we have now calculated both sides of~\eqref{M1_equality}, meaning that we now have a value for the first moment of time of the conditioned random walk, given by 
\begin{equation*}\label{Moment1}
\mu_1 = \frac{N^2 - 1}{3}.
\end{equation*}

The procedure to find the next two moments is not too substantially different: just repeat the same steps of verification and evaluation on $M_{n}^{(1)}$'s siblings
\begin{alignat*}{1}
& M_{n}^{(2)} = X_{n}^5 - 10nX_{n}^3 +(15n^2 +10n)X_n \\
& M_{n}^{(3)} = X_{n}^7 - 21n X_{n}^{5} +(105n^2 + 70n)X_{n}^3 - (105n^3+210n^2+112n)X_n.
\end{alignat*}
These two martingales reveal the next two moments, which are given by
\begin{alignat*}{1}
& \mu_2 = \frac{7N^4 -20N^2+13}{45} \\ 
& \mu_3 = \frac{31 N^6 - 147 N^4 + 189 N^2 - 73}{315} . 
\end{alignat*}
This is all we need to compute the asymptotic skew of the conditioned fixation times. In the limit of large $N$, only the dominant terms of each $\mu_i$ will survive.  The $N$'s cancel out in this limit, leading to a constant given by
\begin{equation*}
\mbox{Skew} = \frac{\mu_3 - 3\mu_1 \mu_2 + 2 \mu_{1}^3}{\left( \mu_2 - \mu_{1}^2 \right)^{3/2}} \approx \frac{8}{7}\left(\frac{5}{2} \right)^{1/2} \approx 1.807. 
\end{equation*}

The conclusion is that in the limit of large $N$, the distribution of fixation times for a conditioned random walk will always be positive. Therefore, we expect right-skewed distributions to be typical for the $r=1$ limit of Birth-death dynamics.

\clearpage

\bibliography{Network_Evo_Bib}

\clearpage

\section*{Acknowledgments}
We thank David Aldous, Rick Durrett, Remco van der Hofstad, Lionel Levine, Piet Van Mieghem, and Steve Schiff for comments. This research was supported by a Sloan Fellowship and NSF Graduate Research Fellowship grant DGE-1650441 to Bertrand Ottino-L\"{o}ffler in the Center for Applied Mathematics at Cornell, and by NSF grants DMS-1513179 and CCF-1522054 to Steven Strogatz. Jacob Scott acknowledges the NIH for their loan repayment grant. 

\section*{Author Contributions}  
B.O.L. provided mathematical analysis and simulations, J.G.S. provided biomedical background, and S.H.S. provided mathematical background. All authors designed the study and wrote the manuscript. 

\section*{Author Information}
The authors declare no competing financial interests. Correspondence and requests for materials should be addressed to
S.H.S. (strogatz@cornell.edu).

\end{document}